\documentclass[aps,pra]{revtex4-2}
\usepackage{amsmath}
\usepackage{amsfonts}
\usepackage{amssymb}
\usepackage{bm}
\usepackage{graphicx}
\begin{document}

\title{Generic equations for long gravity waves in incompressible fluid with finite amplitude}
\affiliation{Centre for Engineering Quantum Systems, School of Mathematics and Physics,
The University of Queensland, Brisbane, Queensland 4072, Australia}
\author{ Vladimir I. Kruglov}

\affiliation{Centre for Engineering Quantum Systems, School of Mathematics and Physics, The University of Queensland, Brisbane, Queensland 4072, Australia}


\begin{abstract}
	
We present the derivation of generic equations describing the long gravity waves in incompressible fluid with decaying effect. We show that in this theory the only restriction to the surface deviation is connected with the stability condition for  the waves. Derivation of these generic equations is based on Euler equations for inviscid incompressible fluid and definition of dynamic pressure which leads to correct dispersion equation for gravity waves. These derived generic equations for velocity of fluid and the surface deviation describe the propagation of long gravity waves in incompressible fluid with finite amplitude. We also have found the necessary and sufficient conditions for generic equations with dissipation of energy or decaying effect. The developed approach can significantly improve the accuracy of theory for long gravity waves in incompressible fluid. We also have found the quasi-periodic and solitary wave solutions for generic equations with decaying effect. 

\end{abstract}

\maketitle

\section{Introduction}

The Korteweg-de Vries equation (KdV) is successful physical equation
describing the shallow water waves with small finite amplitude \cite{1,2, 3,4}. The KdV equation also describes pressure waves in a bubble-liquid mixture \cite{5}; acoustic waves and heat pulses in anharmonic crystals \cite{6,7,8}; magnetic-sonic waves in magnetic plasma \cite{9,10,11,12}; electron plasma waves in a cylindrical plasma \cite{13,14}; and ion acoustic waves \cite{15,16,17,18}. Other impotent results for the Korteweg-de Vries equation have also presented for an example in Refs. \cite{28,29,30,31}.
The KdV equation is tested experimentally as a model for 
moderate amplitude waves propagating in one direction in relatively shallow water of uniform depth. For a wide range of initial data, comparisons are made between the asymptotic wave forms observed and those predicted by the theory in terms of the number of solitons that evolve, the amplitude of the leading soliton, the asymptotic shape of the wave and other qualitative features \cite{32}. Computations made in this work by Hammack and Segur suggest that the KdV equation predicts the amplitude of leading soliton to within the expected error due to viscosity (12$\%$) when the non-decayed amplitudes are less than about a quarter of the water depth. The agreement to within about 20$\%$ is observed over the entire range of experiments examined, including those with initial data for which the non-decayed amplitudes of the leading soliton exceed half the fluid depth.
 
The derivation of weakly nonlinear dispersive wave equations, such as the KdV equation is based on an asymptotic expansion of the water wave equations in the small parameters of the wave amplitude to depth ratio and the depth to wavelength ratio. However, it has been found that higher order terms in this asymptotic expansion are needed to adequately model physical waves \cite{33}.
The extended KdV equation describing the long gravity waves without
strong limitation to surface deviation is derived in Ref. \cite{34}. 
The derivation of this extended KdV equation is based on
Euler equations for incompressible fluid. This generalized
extended KdV equation also describes the decaying effect of the waves. 

In this paper we present the derivation of other generic equations for the long waves or shallow fluid. We not assume in this theory a small wave amplitude condition  $\vert\eta\vert/h_{0}\ll 1$ where $\eta(x,t)$  is the surface deviation of the waves under an equilibrium level $h_{0}$.  The only limitation for wave amplitude in the generic equations is connected with the stability condition for the gravity waves. We also have found the necessary and sufficient conditions for generic equations with dissipation of energy or decaying effect.
The results in this paper are presented as follows. Sec. II presents the derivation of system  equations describing the gravity waves in incompressible fluid with  decaying effect. In Sec. III we derive the generic equations for long gravity waves with finite amplitude which significantly simplify the theory developed in Sec. II.
In Sec. IV and V, we consider the propagation of traveling gravity waves in shallow water. Finally, we summarize the results in Sec. VI. We also present important results in Appendixes A, B and C.

\section{Gravity waves in incompressible fluid with decaying effect}

In general case the gravity waves in shallow water of uniform depth can be described by incompressible Euler equations with additional term connected to decaying effect:
\begin{equation}
\partial_{t}u+u\partial_{x}u+w\partial_{z}u=-\frac{1}{\rho}\partial_{x}P-\Gamma u,
\label{1}
\end{equation}%
\begin{equation}
\partial_{t}w+u\partial_{x}w+w\partial_{z}w=-\frac{1}{\rho}\partial_{z}P-g,
\label{2}
\end{equation}%
\begin{equation}
\partial_{x}u+\partial_{z}w=0,
\label{3}
\end{equation}
where $\mathbf{v}=(u, 0, w)$ is the velocity, $P$ is pressure, $g$ is the acceleration by gravity, and $\rho=const$. The term $-\Gamma u$ in Eq. (\ref{1}) is connected with decaying effect of gravity waves. The water depth $h(x,t)$ for waves propagating to x-direction depends on the time $t$ and longitudinal coordinate $x$. We define the water depth for the waves as $h(x,t)=h_{0}+\eta(x,t)$ where $\eta(x,t)$ is the surface deviation under the equilibrium level $h_{0}$. We also define the small parameter in the theory as $\epsilon^{2}\ll 1$ where $\epsilon=h_{0}/l$  and $l$ is the characteristic length of the gravity wave. This small parameter $\epsilon^{2}$ means that we consider the propagation of long gravity waves. 

It is shown in the Appendix A that the full pressure $P$ can be presented as the sum of static $P_{g}$ and dynamic $P_{d}$ gravitational pressures respectively. The static pressure is given by equation as $\rho^{-1}\partial_{z}P_{g}=-g$. 
The full pressure $P$ and the static pressure $P_{g}$ are given by
\begin{equation}
P=P_{g}+P_{d},~~~~P_{g}=P_{0}+\rho g[h(x,t)-z],
\label{4}
\end{equation}%
where $z$ is the vertical coordinate and $P_{0}$ is the pressure at $z=h$. The term  $\rho^{-1}\partial_{x}P$ can be written by Eq. (\ref{4}) as
\begin{equation}
\frac{1}{\rho}\partial_{x}P=g\partial_{x}\eta+\frac{1}{\rho}\partial_{x}P_{d},
\label{5}
\end{equation}
where the dynamic pressure $P_{d}(x,t)$ depends on variables $x$ and $t$. We also use the standard assumption that the velocity $u=u(x,t)$ depends on variables $x$ and $t$ only. This is correct when the initial velocity $u(x,0)$ does not depend on variable $z$. In this case Eqs. (\ref{1}) and (\ref{5}) lead to the equation,
\begin{equation}
\partial_{t}u+u\partial_{x}u+g\partial_{x}\eta+\mathcal{G}=
-\Gamma u,
\label{6}
\end{equation}%
where the function $\mathcal{G}$ is given by
\begin{equation}
\mathcal{G}(x,t)\equiv\rho^{-1}\partial_{x}P_{d}(x,t).	
\label{7}
\end{equation}
The term $\mathcal{G}$ in this equation is necessary for correct description of the dispersion relation in the first order to small parameter $\epsilon^{2}$. Thus, the introduction of dynamic gravitational pressure $P_{d}$ in Eqs. (\ref{4}) and (\ref{5}) leads to Eq. (\ref{6}) which also describes the decaying effect for propagating waves. 

Now we consider the derivation of conservation equation in a proper form.
Integration of the conservation Eq. (\ref{3}) yields
\begin{equation}
\int_{0}^{h}(\partial_{x}u+\partial_{z}w)dz=0.
\label{8}
\end{equation}%
We have the apparent boundary conditions as $[w]_{z=0}=0$ and $[w]_{z=h}=0$. Thus, Eq. (\ref{8}) can be written as
\begin{equation}
\partial_{x}\int_{0}^{h}udz-[u]_{z=h}\partial_{x}h=0.
\label{9}
\end{equation}%
Considering the boundary condition $Dh/Dt=\partial_{t}h+[u]_{z=h}\partial_{x}h=0$ and the velocity $u=u(x,t)$ which does not depend on variable $z$ we have by Eq. (\ref{9}) the following conservation equation,
\begin{equation}
\partial_{t}h+\partial_{x}(uh)=0.
\label{10}
\end{equation}%
Thus, the incompressible Euler equations with conservation equation lead for long gravity waves to the system of Eqs. (\ref{6}) and (\ref{10}). It is show below  (see also Appendix B) that we can choose 
the function $\mathcal{G}(x,t)$ in the following form:
\begin{equation}
\mathcal{G}(x,t)=-\beta\partial_{x}^{2}\partial_{t}u.	
\label{11}
\end{equation}
This function leads to correct dispersion equation for waves on water surface  in the first order to small parameter $\epsilon^{2}$. We have found this explicit linear form of function $\mathcal{G}$ using the following two conditions: the function $\mathcal{G}$ depends only on the first order of time derivative to velocity $u$, the function $\mathcal{G}$ yields the correct dispersion equation for waves on water surface in the long wave approximation. 
We also have shown in Appendix B that the function in Eq. (\ref{11})
leads to the same dispersion relation for the system of Eqs. (\ref{6}) and (\ref{10}) as the Boussinesq equations.
Using this function $\mathcal{G}(x,t)$ and Eq. (\ref{6}) we have the following equation:
\begin{equation}
\partial_{t}u+u\partial_{x}u+g\partial_{x}h-\beta\partial_{x}^{2}\partial_{t}u=-\Gamma u.
\label{12}
\end{equation}%
Thus, we have derived the closed system of Eqs. (\ref{10}) and (\ref{12}) for the functions $h(x,t)$ and $u(x,t)$. The parameter $\beta$ is found below by condition that the system of Eqs. (\ref{10}) and (\ref{12}) yields the correct dispersion relation in the first order to small parameter $\epsilon^{2}$.

The linearized system of Eqs. (\ref{10}) and (\ref{12}) without term connected with decaying effect is given by
\begin{equation}
\partial_{t}\eta+h_{0}\partial_{x}(u)=0,
\label{13}
\end{equation}%
\begin{equation}
\partial_{t}u+g\partial_{x}\eta-\beta\partial_{x}^{2}
\partial_{t}u=0.
\label{14}
\end{equation}%
The substitution of plain waves:
\begin{equation}
\eta=A\exp[i(\kappa x-\omega t)],~~~~u=B\exp[i(\kappa x-\omega t)], \label{15}
\end{equation}%
to a system of Eqs. (\ref{13}) and (\ref{14}) yields the equation $B=\omega A/\kappa h_{0}$, and the following dispersion relation, 
\begin{equation}
\omega^{2}=\frac{c_{0}^{2}\kappa^{2}}{1+\beta\kappa^{2}},
\label{16}
\end{equation}
where $\kappa$ is the wave number and $c_{0}=\sqrt{gh_{0}}$ is the characteristic velocity. This
characteristic velocity $c_{0}$ is connected with dispersion equation for the waves on water surface.
The dispersion relation for waves on liquid surface \cite{35} is
\begin{equation}
\omega^{2}=\left(1+\frac{\gamma\kappa^{2}}{\rho g}\right) g\kappa\tanh(\kappa h_{0})=\left(1+\frac{\gamma\kappa^{2}}{\rho g}\right) g\kappa\left[\kappa h_{0}-\frac{1}{3}(\kappa h_{0})^{3}+\frac{2}{5}(\kappa h_{0})^{5}-... \right],
\label{17}
\end{equation}%
where $\gamma$ is the surface tension. In the case when $\kappa^{2} h_{0}^{2}\ll 1$
the decomposition presented in Eq. (\ref{17}) can be written as
 \begin{equation}
\omega^{2}=c_{0}^{2}\kappa^{2}-c_{0}^{2}\left(\frac{h_{0}^{2}}{3}-\frac{\gamma}{\rho g}\right)\kappa^{4}+...~ .
\label{18}
\end{equation}%
We can also present Eq. (\ref{16}) as the following  decomposition,
\begin{equation}
\omega^{2}=c_{0}^{2}\kappa^{2}-\beta c_{0}^{2}\kappa^{4}+... ~.
\label{19}
\end{equation}%
The requirement that Eqs. (\ref{18}) and (\ref{19}) yield the same dispersion equation for long gravity waves leads to parameter $\beta$ as
\begin{equation}
\beta=\frac{h_{0}^{2}}{3}-\frac{\gamma}{\rho g}.	
\label{20}
\end{equation}
We emphasize that the Boussinesq equations (see Appendix B) have the same dispersion relation as in Eq. (\ref{16}) with $\gamma=0$. Nevertheless the Boussinesq equation (\ref{7b}) considerably differs from Eq. (\ref{12}) derived in this paper. Moreover, Eqs. (\ref{7}) and (\ref{11}) directly lead to the following dynamic pressure:
\begin{equation}
P_{d}=-\beta\rho\partial_{x}\partial_{t}u,	
\label{21}
\end{equation}
where parameter $\beta$ is given in Eq. (\ref{20}).

\section{Generic equations for long gravity waves with finite amplitude}

The system of Eqs. (\ref{10}) and (\ref{12}) derived in Sec. II can be reduced to more simple equations which we call as generic equations for long gravity waves. The derivation of reduced generic equations for long gravity waves is based on the following transformation:
\begin{equation}
u(x,t)=2\sqrt{gh_{0}+g\eta(x,t)-f(x,t)}-2\sqrt{gh_{0}},
\label{22}
\end{equation}%
where $f(x,t)$ some new function which we define below. This transformation means that the velocity $u(x,t)$ depends on two functions as $\eta(x,t)$ and $f(x,t)$.
We emphasize that transformation defined in Eq. (\ref{22}) with $f(x,t)\equiv 0$ is a Riemann invariant for the system of Eqs. (\ref{6}) and (\ref{10}) when $\mathcal{G}(x,t)\equiv 0$ and $\Gamma=0$. The transformation given in Eq. (\ref{22}) can also be written as
\begin{equation}
\eta(x,t)=\frac{1}{g}f(x,t)+\frac{c_{0}}{g}u(x,t)
+\frac{1}{4g}u^{2}(x,t),
\label{23}
\end{equation}
which leads to equation,
\begin{equation}
g\partial_{x}\eta=\partial_{x}f+c_{0}\partial_{x}u
+\frac{1}{2}u\partial_{x}u.
\label{24}
\end{equation}
The system of Eqs. (\ref{12}) and (\ref{24}) yield the following equation,
\begin{equation}
\partial_{t}u+c_{0}\partial_{x}u
+\frac{3}{2}u\partial_{x}u+\partial_{x}f-\beta\partial_{x}^{2}\partial_{t}u=-\Gamma u.
\label{25}
\end{equation}
We define the unknown function $f(x,t)$ by equation
\begin{equation}
\partial_{x}f=\alpha\partial_{x}^{2}\partial_{t}u,
\label{26}
\end{equation}%
which transforms Eq. (\ref{25}) to the following form:
\begin{equation}
\partial_{t}u+c_{0}\partial_{x}u
+\frac{3}{2}u\partial_{x}u-\sigma\partial_{x}^{2}\partial_{t}u=-\Gamma u,
\label{27}
\end{equation}%
with $\sigma =\beta-\alpha$. The linearized Eq. (\ref{27}) without dissipation therm (for $\Gamma=0$) is given by
\begin{equation}
\partial_{t}u+c_{0}\partial_{x}u-\sigma\partial_{x}^{2}
\partial_{t}u=0.
\label{28}
\end{equation}%
This linear equation has the following dispersion relation, 
\begin{equation}
\omega=\frac{c_{0}\kappa}{1+\sigma\kappa^{2}}=c_{0}\kappa(1-\sigma\kappa^{2}+...).
\label{29}
\end{equation}
The dispersion relation given by Eq. (\ref{19}) can be written as
\begin{equation}
\omega=\pm(c_{0}^{2}\kappa^{2}-\beta c_{0}^{2}\kappa^{4}+...)^{1/2}
=\pm c_{0}\kappa\left(1-\frac{1}{2}\beta\kappa^{2}-...\right),
\label{30}
\end{equation}
which also follows from Eq. (\ref{16}). We emphasize that the closed system of Eqs. (\ref{10}) and (\ref{12}) describes the waves which can propagate to the left and to the right simultaneously. However, Eq. (\ref{27}) consider the waves propagating to the right only. Hence, we should choose the positive brunch in the dispersion relation given in Eq. (\ref{30}) which together with Eq. (\ref{29}) yield the following equations: $\sigma=\beta/2$ and $\alpha=\beta-\sigma=\sigma$. 
Thus, Eqs. (\ref{26}) and (\ref{20}) with above relations lead to the following equations,
\begin{equation}
f=\sigma\partial_{x}\partial_{t}u,~~~~\sigma=\frac{h_{0}^{2}}{6}-\frac{\gamma}{2\rho g}.
\label{31}
\end{equation}%
The surface deviation given in Eq. (\ref{23}) and Eq. (\ref{31}) lead to equation for the function $\eta(x,t)$ as
\begin{equation}
\eta(x,t)=\frac{1}{g}\sigma\partial_{x}\partial_{t}u+\frac{c_{0}}{g}u(x,t)
+\frac{1}{4g}u^{2}(x,t).
\label{32}
\end{equation}%
We can defined the following dimensionless variables $\tau=c_{0}t/l$, $\lambda=x/l$ and the functions $\tilde{\eta}$ and $\tilde{u}$ as
\begin{equation}
\tilde{\eta}=\frac{\eta}{h_{0}},~~~~\tilde{u}=\frac{u}{c_{0}}.
\label{33}
\end{equation}
These dimensionless variables and functions lead to dimensionless form of Eq. (\ref{32}) as
\begin{equation}
\tilde{\eta}=\frac{\chi}{6}\epsilon^{2}\partial_{\lambda}
\partial_{\tau}\tilde{u}+\tilde{u}+\frac{1}{4}\tilde{u}^{2},
\label{34}
 \end{equation}%
where $\chi=1-3\gamma/\rho gh_{0}^{2}$ and $\epsilon=h_{0}/l$.
The long gravity waves approximation ($\epsilon^{2}\ll 1$) yields the equation:
\begin{equation}
\tilde{\eta}=\tilde{u}+\frac{1}{4}\tilde{u}^{2}.
\label{35}
\end{equation}
Hence, the system of Eqs. (\ref{27}) and (\ref{35}) can be written in the form:
\begin{equation}
\partial_{t}u+c_{0}\partial_{x}u
+\frac{3}{2}u\partial_{x}u-\sigma\partial_{x}^{2}\partial_{t}u=
-\Gamma u,
\label{36}
\end{equation}%
\begin{equation}
\eta(x,t)=\frac{c_{0}}{g}u(x,t)+\frac{1}{4g}u^{2}(x,t).
\label{37}
\end{equation}

Note that we also have the conservation Eq. (\ref{10})
which should be consistent with the system of Eqs. (\ref{36}) and (\ref{37}). In  Appendix C we show that Eq. (\ref{10}) is satisfied for two condition: $\epsilon^{2}\ll 1$ and $6\Gamma h_{0}/c_{0}\chi\ll 1$. Moreover, these two conditions are necessary and sufficient for the theory based on equations (\ref{36}) and (\ref{37}). In the case when dissipation term is negligible small we have for generic Eqs. (\ref{36}) and (\ref{37}) only the long wave condition as $\epsilon^{2}\ll 1$.
We note that the system of Eqs. (\ref{36}) and (\ref{37}) can be written in other equivalent form using the function $\zeta(x,t)$:
\begin{equation}
\zeta(x,t)=\frac{c_{0}}{g}u(x,t).
\label{38}
\end{equation}
The system of Eqs. (\ref{36}) and (\ref{37}) with definition (\ref{38}) has the form,
\begin{equation}
\partial_{t}\zeta+c_{0}\partial_{x}\zeta
+\mu\zeta\partial_{x}\zeta-\sigma\partial_{x}^{2}\partial_{t}
\zeta=-\Gamma \zeta,
\label{39}
\end{equation}%
\begin{equation}
\eta(x,t)=\zeta(x,t)+\frac{1}{4h_{0}}\zeta^{2}(x,t),
\label{40}
\end{equation}
where the parameters $\mu$ and $\sigma$ are
\begin{equation}
\mu=\frac{3c_{0}}{2h_{0}},~~~~\sigma=\frac{\chi h_{0}^{2}}{6},~~~~\chi=1-\frac{3\gamma}{\rho gh_{0}^{2}}.
\label{41}
\end{equation}%
The experimental observations show that the solitary waves propagating in shallow water are stable when the condition $\eta (x,t)<\eta_{0}$ is satisfied where $\eta_{0}/h_{0}\approx  0.7$ \cite{35}. Our theoretical stability condition leads to relation $\eta_{0}/h_{0}\approx 0.76$ which is close to the experimental observations. It is shown in Appendix C that these generic equation are satisfied for long waves approximation $\epsilon^{2}\ll 1$ and additional condition as $6\Gamma c_{0}/g\chi\ll 1$. Thus, in the case when we can neglect the decaying effect the generic equations are satisfied for long wave approximation or shallow water condition $\epsilon^{2}\ll 1$.

\section{Traveling gravity waves }

In this section we consider the propagation of traveling gravity waves in shallow water without decaying effect. In this case the only limitation to surface deviation is connected with stability condition for the gravity waves. 
The term describing dissipation of energy can be neglected for propagating distances $L$
satisfying the condition $\Gamma L/v_{0}\ll 1$ where $v_{0}$ is the velocity of gravity waves. In this case  Eqs. (\ref{39}) and (\ref{40}) with $\Gamma=0$ have the form:
\begin{equation}
\partial_{t}\zeta+c_{0}\partial_{x}\zeta
+\mu\zeta\partial_{x}\zeta-\sigma\partial_{x}^{2}\partial_{t}
\zeta=0,
\label{42}
\end{equation}%
\begin{equation}
\eta(x,t)=\zeta(x,t)+\frac{1}{4h_{0}}\zeta^{2}(x,t).
\label{43}
\end{equation}
Integration of Eq. (\ref{42}) leads to the second order nonlinear differential equation. This equation for the function $\zeta(x,t)=U(s)$ has the form:
\begin{equation} 
\frac{d^{2}U}{ds^{2}}+aU^{2}+bU+C_{1}=0,
\label{44}
\end{equation}
where  $s=x-v_{0}t$ and $C_{1}$ is integration constant. The parameters $a$ and $b$ are
\begin{equation} 
a=\frac{\mu}{2\sigma v_{0}},~~~~b=\frac{c_{0}-v_{0}}{\sigma v_{0}}.	
\label{45}
\end{equation}
The elliptic differential equation (\ref{44}) yields the periodic solution as
\begin{equation} 
\zeta(x,t)=A_{0} k^{2}\mathrm{cn}^{2}(W_{0}\xi,k),
\label{46}
\end{equation}
where $A_{0}$ is an arbitrary positive constant, $\xi=x-x_{0}-v_{0}t$, and $\mathrm{cn}(z,k)$ is the elliptic Jacobi function. The parameters $W_{0}$ and $v_{0}$ in this periodic solution are
\begin{equation} 
W_{0}=\frac{1}{2h_{0}}\sqrt{\frac{3A_{0} c_{0}}{\chi h_{0}v_{0}}},~~~~
v_{0}=c_{0}+\frac{c_{0}A_{0}}{2h_{0}}(2k^{2}-1).	
\label{47}
\end{equation}
This periodic solution depends for two positive free parameters as $0< k< 1$ and $A_{0}>0$. Equations (\ref{43}) and (\ref{46}) lead to solution for the function $\eta(x,t)$ as
\begin{equation} 
\eta(x,t)=A_{0} k^{2}\mathrm{cn}^{2}(W_{0}\xi,k)
+\frac{A_{0}^{2}k^{4}}{4h_{0}}\mathrm{cn}^{4}(W_{0}\xi,k).
\label{48}
\end{equation}
The periodic solution in Eq. (\ref{48}) reduces to the solitary wave for limiting case with $k=1$ as 
\begin{equation} 
\eta(x,t)=A_{0} \mathrm{sech}^{2}(W_{0}\xi)+\frac{A_{0} ^{2}}{4h_{0}}\mathrm{sech}^{4}(W_{0}\xi).
\label{49}
\end{equation}
The inverse width $W_{0}$ and velocity $v_{0}$ in this soliton solution are 
\begin{equation} 
W_{0}=\frac{1}{2h_{0}}\sqrt{\frac{6A_{0}}{\chi (A_{0}+2h_{0})}},~~~~
v_{0}=c_{0}+\frac{c_{0}A_{0}}{2h_{0}}.	
\label{50}
\end{equation}
The periodic solution given in Eq. (\ref{48}) for small parameter $k$ ($k\ll 1$) has the form,
\begin{equation} 
\eta(x,t)=A_{0} k^{2}\cos^{2}(W_{0}\xi)+\frac{A_{0}^{2}k^{4}}{4h_{0}}
\cos^{4}(W_{0}\xi),
\label{51}
\end{equation}
where parameters $W_{0}$ and $v_{0}$ are given in Eq. (\ref{47}).

\section{Decaying wave solutions for generic equations}

In this section we consider the propagation of gravity waves in shallow water with decaying effect. We have shown (see also Appendix C) that when two conditions $\epsilon^{2}\ll 1$ and $\Gamma\ll \chi c_{0}/6h_{0}$ are satisfied the generic equations (\ref{39}) and (\ref{40}) describe the decaying long gravity waves in incompressible fluid with finite amplitude. Using the method developed in the Ref. \cite{34} we can show that decaying wave solutions depend on amplitude  
$A(t)$ which satisfies to differential equation:
\begin{equation} 
\frac{dA(t)}{dt}+\Gamma A(t)=0.
\label{52}
\end{equation}
Thus, the time dependent amplitude $A(t)$ is given by
\begin{equation} 
A(t)=A_{0}\exp(-\Gamma t),
\label{53}
\end{equation}
where $A_{0}$ is an arbitrary positive constant.
The decaying quasi-periodic solution of Eqs. (\ref{39}) and (\ref{40}) for the function $\zeta(x,t)$ is
\begin{equation} 
\zeta(x,t)=A(t) k^{2}\mathrm{cn}^{2}(W(t)S(t),k),
\label{54}
\end{equation}
where $S(t)=x-x_{0}-\int_{0}^{t}v(t^{\prime})dt^{\prime}$. The inverse width $W(t)$ and velocity $v(t)$ in this quasi-periodic solution are
\begin{equation} 
W(t)=\frac{1}{2h_{0}}\sqrt{\frac{3A(t) c_{0}}{\chi h_{0}v(t)}},~~~~
v(t)=c_{0}+\frac{c_{0}A(t)}{2h_{0}}(2k^{2}-1).	
\label{55}
\end{equation}
This quasi-periodic solution depends on two positive free parameters as $0< k< 1$ and $A_{0}>0$. Using Eq. (\ref{55}) we
have the variable $S(t)$ in the following explicit form, 
\begin{equation} 
S(t)=x-x_{0}-c_{0}t-\frac{c_{0}A_{0}}{2h_{0}\Gamma}(2k^{2}-1)
[1-\exp(-\Gamma t)].	
\label{56}
\end{equation}
Equations (\ref{40}) and (\ref{54}) lead to solution for the function $\eta(x,t)$ as
\begin{equation} 
\eta(x,t)=A(t) k^{2}\mathrm{cn}^{2}(W(t)S(t),k)
+\frac{A^{2}(t)k^{4}}{4h_{0}}\mathrm{cn}^{4}(W(t)S(t),k).
\label{57}
\end{equation}
The quasi-periodic solution given in Eq. (\ref{57}) reduces to the solitary wave for limiting case with $k=1$ as 
\begin{equation} 
\eta(x,t)=A(t) \mathrm{sech}^{2}(W(t)S(t))+\frac{A ^{2}(t)}{4h_{0}}\mathrm{sech}^{4}(W(t)S(t)),
\label{58}
\end{equation}
where the inverse width $W(t)$ and velocity $v(t)$ are
\begin{equation} 
W(t)=\frac{1}{2h_{0}}\sqrt{\frac{3A(t) c_{0}}{\chi h_{0}v(t)}},~~~~
v(t)=c_{0}+\frac{c_{0}A(t)}{2h_{0}},	
\label{59}
\end{equation}
and the function $S(t)$ in this soliton solution is
\begin{equation} 
S(t)=x-x_{0}-c_{0}t-\frac{c_{0}A_{0}}{2h_{0}\Gamma}
[1-\exp(-\Gamma t)].	
\label{60}
\end{equation}
The periodic solution in Eq. (\ref{57}) for small parameter $k$ ($k\ll 1$) has the form,
\begin{equation} 
\eta(x,t)=A(t) k^{2}\cos^{2}(W(t)S(t))+\frac{A^{2}(t)k^{4}}{4h_{0}}
\cos^{4}(W(t)S(t)),
\label{61}
\end{equation}
where the functions $W(t)$, $v(t)$ and $S(t)$ are given in Eqs. (\ref{55}) and (\ref{56}) with small parameter $k\ll 1$.

\section{Conclusion}

In this paper, we have derived the generic equations for the water waves with arbitrary amplitudes. The only restriction for the surface deviation connected to the gravity waves is the stability condition for the waves. We use in this derivation of generic equations the long-wave approximation. These generic equations
also consider the decaying effect for the propagating waves. This decaying effect  connected to dissipation of energy is important for describing the propagation of waves to long distances. It is shown that the term describing decaying effect  depends on two parameters as the kinematic viscosity $\nu$ and the capillary length $\lambda_{c}$. The explicit form for the decaying term is derived using the dimensional analysis with the critical parameters $\nu$ and $\lambda_{c}$. In the paper of Hammack and Segur \cite{32} is demonstrated  that the agreement to within about 20$\%$ is observed for KdV equation over the entire range of experiments examined for moderate amplitudes of the waves. We emphasize that this difference
between experiments with moderate amplitudes of the waves and theoretical description
takes place because the KdV equation describes only the waves with infinitesimal small amplitudes. It is remarkable that the generic equations presented in this paper describe the stable waves with arbitrary amplitudes.
We also have found the necessary and sufficient conditions for generic equations with dissipation energy for the waves or decaying
effect. The approach based on generic equations can significantly improve the accuracy of theory for long gravity waves in incompressible fluid. In this paper we also have found the quasi-periodic and solitary wave solutions for generic equations with decaying effect. 

\appendix

\section{Static and dynamic gravitational pressures in incompressible fluid}

We have defined the following dimensionless variables:
$\tau=c_{0}t/l$, $\lambda=x/l$ and $\xi=z/l$. The dimensionless velocities $\tilde{u}$ and $\tilde{w}$ are given by relations as $u=c_{0}\tilde{u}$ and $w=\epsilon c_{0}\tilde{w}$ respectively. The conservation equation (\ref{3}) with
these dimensionless variables has the form:
\begin{equation}
\partial_{\lambda}\tilde{u}+\partial_{\xi}\tilde{w}=0.	
\label{1a}
\end{equation}%
The Euler Eq. (\ref{2}) can be written in the standard form as
\begin{equation}
\frac{Dw}{Dt}=-\frac{1}{\rho}\partial_{z}P-g.
\label{2a}
\end{equation}%
Using defined dimensionless variables  we can write this equation in the form,
\begin{equation}
\epsilon^{2}\frac{D\tilde{w}}{D\tau}=-g^{-1}\left(
\frac{1}{\rho}\partial_{z}P+g\right).
\label{3a}
\end{equation}%
Hence, for long waves $(\epsilon^{2}\ll 1)$ Eq. (\ref{3a}) yields
the following equation for pressure $P$:
\begin{equation}
\frac{1}{\rho}\partial_{z}P+g=0.
\label{4a}
\end{equation}
We can present the full pressure $P$ as the sum of two terms:
\begin{equation} 
P=P_{g}+P_{d},
\label{5a}
\end{equation}
where $P_{g}$ and $P_{d}$ are the static and dynamic gravitational pressures. We note that dynamic gravitational pressure $P_{d}$ is necessary for correct description of the dispersion relation in the first order to small parameter $\epsilon^{2}$. The static gravitation pressure $P_{g}$ depends on the liquid depth $h(x,t)$ and the vertical coordinate $z$, and the dynamic gravitational pressure $P_{d}$ depends on variables $x$ and $t$. Thus, Eqs. (\ref{4a}) and (\ref{5a}) yield the equation for static pressure as 
\begin{equation}
\frac{1}{\rho}\partial_{z}P_{g}=-g,
\label{6a}
\end{equation}
because $\partial_{z}P_{d}=0$.
This equation leads to the following static gravitation pressure: 
\begin{equation} 
P_{g}=P_{0}+\rho g[h(x,t)-z],	
\label{7a}
\end{equation}
where $z$ is the vertical coordinate and $P_{0}$ is the pressure at $z=h$. 
It is shown by Eqs. (\ref{20}) and (\ref{21}) that the dynamic pressure $P_{d}$ is given as
\begin{equation}
P_{d}=-\beta\rho\partial_{x}\partial_{t}u,~~~~\beta=
\frac{h_{0}^{2}}{3}-\frac{\gamma}{\rho g}.	
\label{8a}
\end{equation}

\section{Decaying parameter and dispersion relations}

We accept \cite{34} that the parameter $\Gamma$ depends on kinematic viscosity $\nu$ and the capillary
length $\lambda_{c}$ defined as
\begin{equation}
\lambda_{c}=\sqrt{\frac{\gamma}{g\rho}},	
\label{1b}
\end{equation}%
where $\gamma$ and $\rho$ are the surface tension and mass density respectively. The dimensional analysis with these two
parameters yields 
\begin{equation}
\Gamma=\frac{Q\nu}{\lambda_{c}^{2}}=Q\nu\rho g\gamma^{-1},	
\label{2b}
\end{equation}%
where $Q=\cal{O}$$(1)$ is dimensionless function of temperature. 
The parameters for water are $\nu=0.01\mathrm{cm^{2}/s}$ and $\lambda_{c}=0.276\mathrm{cm}$ (for temperature $T=20^{o}\mathrm{C}$) which yields $\Gamma=0.131\times Q~\mathrm{s^{-1}}$.

The system of Eqs. (\ref{10}) and (\ref{12}) without decaying term is given as
\begin{equation}
\partial_{t}h+\partial_{x}(uh)=0,
\label{3b}
\end{equation}%
\begin{equation}
\partial_{t}u+u\partial_{x}u+g\partial_{x}h-\beta\partial_{x}^{2}\partial_{t}u=0.
\label{4b}
\end{equation}%
This system of equations yields the following dispersion relation,
\begin{equation}
\omega^{2}
=\frac{c_{0}^{2}\kappa^{2}}{1+\beta\kappa^{2}},~~~~\beta=\frac{h_{0}^{2}}{3}-\frac{\gamma}{\rho g}.	
\label{5b}
\end{equation}%
The system of Boussinesq equations \cite{35} has the form,
\begin{equation}
\partial_{t}h+\partial_{x}(uh)=0,
\label{6b}
\end{equation}%
\begin{equation}
\partial_{t}u+u\partial_{x}u+g\partial_{x}h+\frac{1}{3}h_{0}
\partial_{x}\partial_{t}^{2}h=0,
\label{7b}
\end{equation}%
which leads to the dispersion relation as
\begin{equation}
\omega^{2}=\frac{c_{0}^{2}\kappa^{2}}{1+\kappa^{2}h_{0}^{2}/3}.
\label{8b}
\end{equation}%
Thus, these different system of equations have the same dispersion relations in the case when $\gamma=0$.

\section{Conservation equation}

We note that the conservation equation presented in Eq. (\ref{10})
should be consistent with the system of Eqs. (\ref{36}) and (\ref{37}). We have the system of equations:
\begin{equation}
\partial_{t}u+c_{0}\partial_{x}u
+\frac{3}{2}u\partial_{x}u-\sigma\partial_{x}^{2}\partial_{t}u=
-\Gamma u,
\label{1c}
\end{equation}%
\begin{equation}
\eta(x,t)=\frac{c_{0}}{g}u(x,t)+\frac{1}{4g}u^{2}(x,t),
\label{2c}
\end{equation}
and the conservation equation,
\begin{equation}
\partial_{t}h+u\partial_{x}h+h\partial_{x}u=0.
\label{3c}
\end{equation}%
Using Eq. (\ref{2c}) we have the following equations:
\begin{equation}
\partial_{t}h=\frac{c_{0}}{g}\partial_{t}u+\frac{1}{2g}u\partial_{t}u,~~~~\partial_{x}h=\frac{c_{0}}{g}\partial_{x}u+\frac{1}{2g}u\partial_{x}u,
\label{4c}
\end{equation}%
\begin{equation}
h=h_{0}+\eta=\frac{c_{0}^{2}}{g}+\frac{c_{0}}{g}u+\frac{1}{4g}u^{2}.
\label{5c}
\end{equation}%
Substitution of Eqs. (\ref{4c}) and (\ref{5c}) to conservation equation (\ref{3c})
leads to the following equation:
\begin{equation}
c_{0}\partial_{t}u+\frac{1}{2}u\partial_{t}u+2c_{0}u\partial_{x}u+\frac{3}{4}u^{2}\partial_{x}u+c_{0}^{2}\partial_{x}u=0.	
\label{6c}
\end{equation}%
Using the derivative $\partial_{t}u$ in Eq. (\ref{1c}) and
Eq. (\ref{6c}) one can derive the following reduced form of Eq. (\ref{3c}):
\begin{equation}
-\Gamma u+\sigma\partial_{x}^{2}\partial_{t}u=0.
\label{7c}
\end{equation}%
We can also use the dimensionless variables introduced in Appendix A as $\tilde{u}=u/c_{0}$, $\tau=c_{0}t/l$ and $\lambda=x/l$. Thus, Eq. (\ref{7c}) has the following dimensionless form,
\begin{equation}
-\frac{\Gamma c_{0}}{g}\tilde{u}+
\frac{\chi}{6}\epsilon^{3}\partial_{\lambda}^{2}\partial_{\tau}\tilde{u}=0.
\label{8c}
\end{equation}%
This equation is satisfied for long waves approximation and additional  condition as $6\Gamma c_{0}/g\chi\ll 1$. 
Hence, the system of Eqs. (\ref{1c}) and (\ref{2c}) is consistent with conservation equation (\ref{3c}) when the following two conditions are satisfied:
\begin{equation}
\epsilon^{2}\ll 1,~~~~\frac{6\Gamma h_{0}}{c_{0}\chi}\ll 1.
\label{9c}
\end{equation}%
Note that these two conditions are necessary and sufficient for consistency of Eqs. (\ref{1c})-(\ref{3c}).
In the case when decaying effect is negligibly small ($\Gamma\simeq 0$) we have for consistency of Eqs. (\ref{1c})-(\ref{3c}) only the shallow water condition as $\epsilon^{2}\ll 1$.

\end{document}